\documentclass[conference,letterpaper]{IEEEtran}

\usepackage[utf8]{inputenc} 
\usepackage[T1]{fontenc}
\usepackage{url}
\usepackage{ifthen}
\usepackage{cite}
\usepackage[cmex10]{amsmath}   
\usepackage{cases}
\usepackage{multirow}
\usepackage{empheq}
\usepackage{verbatim}
\usepackage{algorithm}
\usepackage{algpseudocode}

\usepackage{siunitx}

\usepackage{kotex}
\usepackage{amssymb}
\usepackage{optidef}
\ifCLASSINFOpdf
  \usepackage[pdftex]{graphicx}
\else
  \usepackage[dvips]{graphicx}
\fi
\usepackage[caption=false,font=footnotesize]{subfig}

\usepackage{color}
\usepackage{multirow}

\newtheorem{proposition}{Proposition}
\newtheorem{remark}{Remark}

\usepackage{xcolor}

\begin{document}
\title{Quantizing for Noisy Flash Memory Channels} 

\author{%
  \IEEEauthorblockN{Juyun Oh\textsuperscript{*}, Taewoo Park\textsuperscript{*}, Jiwoong Im\textsuperscript{*}, Yuval Cassuto\textsuperscript{†}, and Yongjune Kim\textsuperscript{*}}
  \IEEEauthorblockA{\textsuperscript{*}Department of Electrical Engineering, POSTECH, Pohang, South Korea \\
                    Email: \{juyunoh, parktaewoo, jw3562, yongjune\}@postech.ac.kr\\
                    \textsuperscript{†}Viterbi Department of Electrical and Computer Engineering, Technion - Israel Institute of Technology, Haifa, Israel\\
                    Email: ycassuto@ee.technion.ac.il}
  
}

\maketitle

\begin{abstract}

    Flash memory-based processing-in-memory (flash-based PIM) offers high storage capacity and computational efficiency but faces significant reliability challenges due to noise in high-density multi-level cell (MLC) flash memories. 
    Existing verify level optimization methods are designed for general storage scenarios and fail to address the unique requirements of flash-based PIM systems, where metrics such as mean squared error (MSE) and peak signal-to-noise ratio (PSNR) are critical.
    This paper introduces an integrated framework that jointly optimizes quantization and verify levels to minimize the MSE, considering both quantization and flash memory channel errors.
    We develop an iterative algorithm to solve the joint optimization problem.
    Experimental results on quantized images and SwinIR model parameters stored in flash memory show that the proposed method significantly improves the reliability of flash-based PIM systems.      

\end{abstract}

\section{Introduction}

    Processing-in-memory (PIM), also known as in-memory computing, has emerged as a promising alternative to the traditional von Neumann architecture, providing reduced data transfer overhead and improved computational efficiency~\cite{Dupraz2023turning,Kang2020deep,Shanbhag2019shannon}. 
    In particular, flash memory-based PIM (flash-based PIM) offers a significantly larger storage capacity compared to SRAM- and DRAM-based PIM, making it an attractive solution for storing vast datasets and model parameters~\cite{Hasan2020Reliability,Gonugondla2018energy}.
    However, the continued scaling of the flash memory technology introduces significant reliability challenges, particularly for high-density flash memory cells. 
    In $N$-bit per cell multi-level cell (MLC) flash memories, which represent $2^N$ distinct states within a constrained threshold voltage window, noise significantly impairs the accuracy of the resulting computations~\cite{Cai2017error,Kim2022writing}. 

  
    Several techniques have been proposed to minimize the impact of flash-memory channel noise by optimizing the distances between states~\cite{Aslam2016read,Asmani2024write,Kim2012verify}. 
    This optimization problem is commonly referred to as \emph{write voltage} optimization~\cite{Aslam2016read,Asmani2024write} or \emph{verify levels} optimization~\cite{Kim2012verify}, as the distances between states (or write voltages) can be adjusted using the verify levels in incremental step pulse programming (ISPP)~\cite{Suh1995A}.
    However, these approaches are primarily designed for general data-storage scenarios, and thus focus on minimizing the bit error rate (BER), rather than optimizing measures more relevant to PIM such as mean squared error (MSE) or peak signal-to-noise ratio (PSNR). 
    These performance measures fit better the requirements of data-intensive applications in domains such as machine learning and computer vision. They also do not rely on conventional channel coding schemes that introduce high latency in computing environments with stringent latency requirements~\cite{Cilasun2024on}.
    Therefore, new strategies for optimizing verify levels should be developed to ensure the efficient and reliable operation of flash-based PIM systems.


    Optimizing the MSE of the in-memory representation opens the opportunity to optimize jointly with another factor affecting the MSE performance: \emph{quantization}. 
    Since the data stored in memory is itself a quantized version of the application data, the overall MSE performance depends on both the quantization levels (source optimization) and the memory verify levels (channel optimization). 
    Quantization is typically optimized without regards to channel errors, employing classical algorithms such as the Lloyd-Max algorithm~\cite{Lloyd1982least,Max1960quantizing}. 
    The importance of channel errors to the quantization performance has been observed before in~\cite{Kurtenbach1969quantizing}, where the authors introduce the concept of \emph{quantizing for noisy channels}.  
    However, this method assumes fixed channel transition probabilities that cannot be controlled, thus not offering the potential benefits of verify-level optimization that were demonstrated in other applications~\cite{Kim2012verify,Aslam2016read}.
       

    In this paper, we propose an integrated framework that optimizes both quantization levels and verify levels to effectively enhance the reliability of flash-based PIM systems. 
    We formulate a joint optimization problem to minimize the MSE, taking into account errors introduced by quantization and the flash memory channel. 
    In this framework, the optimization variables include the quantization levels and verify levels.
    Since the impacts of these two sets of variables on MSE are interdependent, we develop an iterative algorithm to solve the formulated optimization problem. 
    Unlike conventional approaches that focus separately on verify-level optimization~\cite{Kim2012verify,Aslam2016read,Asmani2024write} and quantization level optimization~\cite{Lloyd1982least,Max1960quantizing,Kurtenbach1969quantizing}, our approach jointly optimizes both the quantization levels and verify levels in an iterative manner.  
    To the best of our knowledge, this is the first work to jointly optimize the quantization and verify levels for flash-based PIM systems.

    To comprehensively evaluate the reliability of data and model parameters stored in the flash-based PIM systems, we conduct two distinct experiments. 
    First, we compare the PSNRs of images stored using the proposed method to that of a conventional approach. 
    Second, we evaluate the PSNRs of the SwinIR model \cite{Liang2021swinir}, assuming its model parameters are quantized to 4 bits and stored in quad-level cell (QLC) flash memory. We then compare the proposed method to the conventional method under these conditions.
    
    The rest of this paper is organized as follows. 
    Section \ref{sec:system} introduces the system model of flash-based PIM and reviews quantization for noisy channels.
    Section \ref{sec:main} derives the MSE for flash-based PIM systems, formulates the joint optimization problem, and presents an iterative algorithm to solve this problem.
    Section \ref{sec:results} provides experimental results, and Section \ref{sec:conclusion} concludes the paper.

    \section{System Model and Channel-Aware Quantization}\label{sec:system}

    In this section, we first provide the system model and then briefly review the scheme of channel-aware quantization from~\cite{Kurtenbach1969quantizing}. 

    \subsection{System Model} \label{subsec:system}

    \begin{figure}[!t]
    \centering
    {\includegraphics[width=0.4\textwidth]{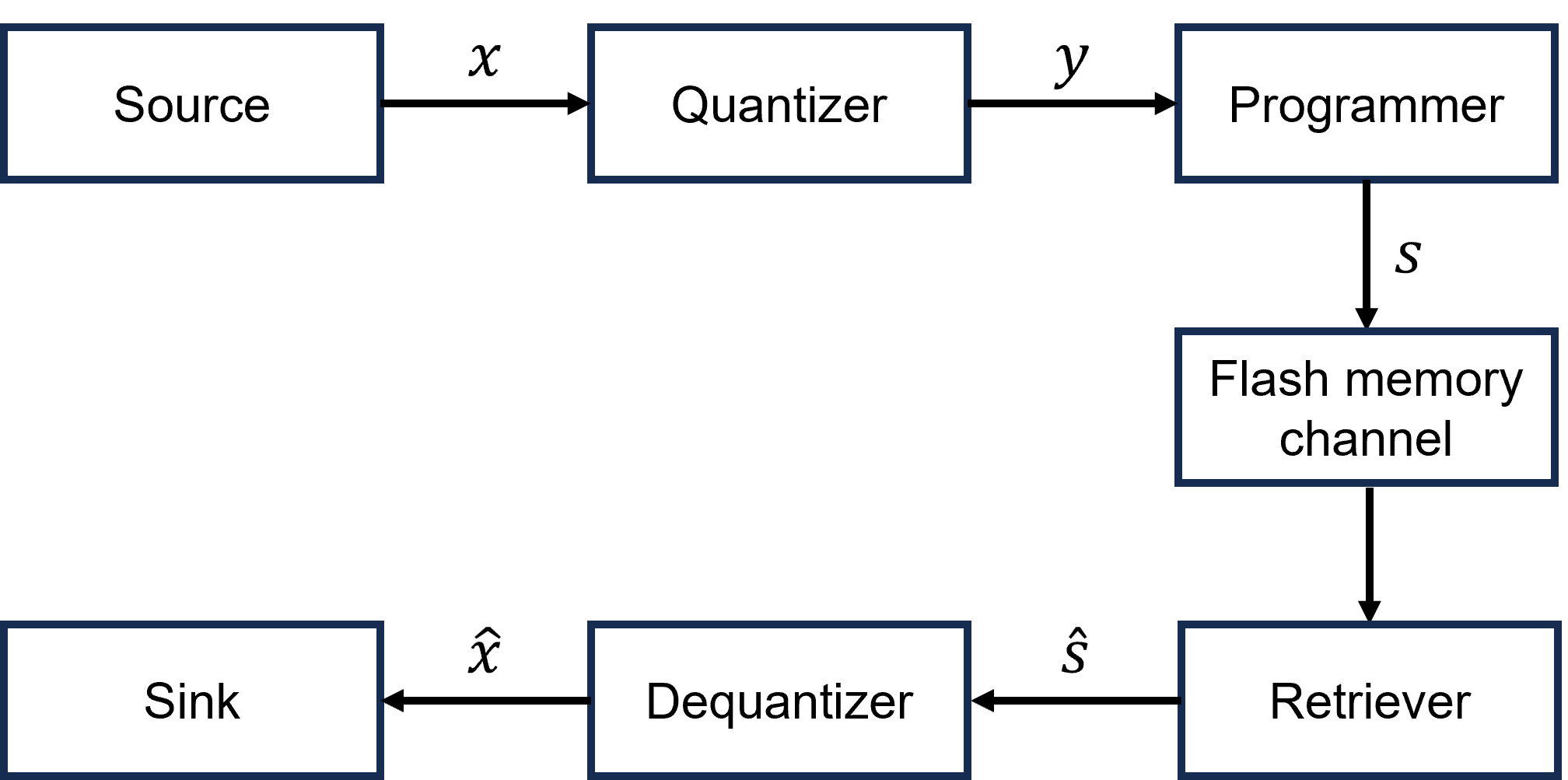}}
        \caption{System model for flash-based PIM systems.}
        \label{fig:model} 
    \vspace{-4mm}
    \end{figure}
    
    The source data is modeled as a realization of a continuous random variable $X$ with the probability density function (PDF) $f_X(x)$. 
    The quantizer maps $x$ into $y$, a value from a finite set of real numbers $\{v_j\}:=\{v_1, \ldots, v_M \}$, following the mapping rule: $y = v_j$ if $u_j < x \le u_{j+1}$. 
    The set $\{u_j\}:= \{u_2, \ldots,u_M\}$ denotes the quantization thresholds, with $u_1$ and $u_{M+1}$ defined as the greatest lower and least upper bound of the input data, respectively.  
    Subsequently, the encoder assigns a binary codeword to the quantizer output $y$, and this codeword is programmed into flash memory cells.
    The output of the noisy memory channel is denoted as $\widehat{s}$. 
    The output of the dequantizer $\widehat{x}$ is the real number $v_j$ mapped back from $\widehat{s}$. 
    The noisy flash memory channel is characterized by a channel transition matrix $P = [P_{i,j}]$, where $P_{i,j} = P(\widehat{x} = v_j | y = v_i) $.  
    The system model is described in Fig.~\ref{fig:model}.
    
    In $N$-bit per cell flash memories, the threshold voltage distribution of memory cells consists of $2^N$ distinct states, ranging from $s_1$ (the erase state) to $s_{2^N}$ (the highest state).    
    This distribution can be approximated as a mixture of Gaussian distributions~\cite{Kim2012verify}.
    Consequently, the threshold voltage distribution $f_T(t)$ is given by
    \begin{align} \label{eq:GMM}
        f_T(t) &= \sum_{i=1}^{2^N} p(s_i)f_i(t) = \sum_{i=1}^{2^N} \frac{p(s_i)}{\sqrt{2\pi}\sigma_i} e^{-\frac{(t-\mu_i)^2}{2\sigma_i^2}},
    \end{align}
    where $t$ denotes the threshold voltage, $f_i(t)$ is the Gaussian PDF with mean $\mu_i$ and variance $\sigma_i^2$ for state $s_i$, and $p(s_i)$ represents the probability of the state $s_i$. 
    
   To simplify the architecture of flash-based PIM systems, we set $M = 2^N$, such that each flash memory cell stores one $N$-bit quantized value. 
    Then, the quantized value $v_i$ is programmed into the state $s_i$ for $i = 1, \ldots, M$. 
    If the retrieved state is $s_j$, then the decoder outputs $\widehat{x} = v_j$. 
    Hence, the channel transition probability is given by
    \begin{equation} \label{eq:channel transition}
    P_{i,j} = P(\widehat{x} = v_j | y = v_i) = P(\widehat{s} = s_j | s = s_i). 
    \end{equation}
    \begin{figure}[!t]
        \centering
        {\includegraphics[width=0.45\textwidth]{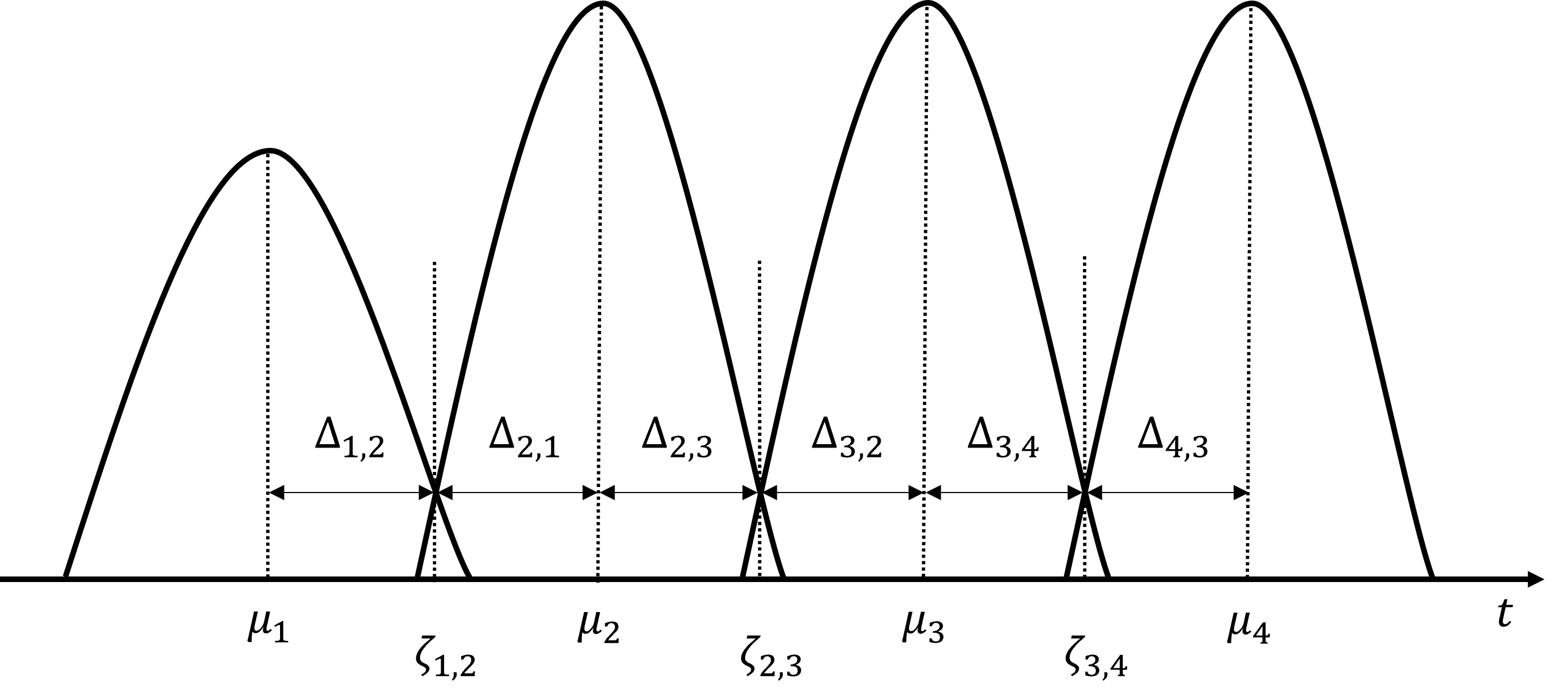}}
            \caption{Threshold voltage distribution for 2-bit per cell flash memories with the constrained window $W = \mu_4 - \mu_1$.}
            \label{fig:distribution} 
        \vspace{-4mm}
        \end{figure}    
    In the Gaussian mixture model, the dominant source of errors arises from the misreading of adjacent states. 
    Hence, flash memory channel errors can be expressed as the following two types of transition probabilities: 
    \begin{align}
        P_{i,i-1}&=P(\widehat{s} = s_{i-1}|s=s_i) \approx Q\left(\frac{\Delta_{i,i-1}}{\sigma_{i}}\right),   \label{eq:transition_prob_1}  \\
        P_{i,i+1}&=P(\widehat{s} = s_{i+1}|s=s_{i}) \approx  Q\left(\frac{\Delta_{i,i+1}}{\sigma_{i}}\right), \label{eq:transition_prob_2}
    \end{align}
    where $\Delta_{i,i-1} = \mu_{i} - \zeta_{i-1,i}$ and $\Delta_{i,i+1} = \zeta_{i,i+1} - \mu_{i} $. 
    Here, $\zeta_{i,i+1}$ is the transition level between $s_{i}$ and $s_{i+1}$.  
    The tail probability function is given by $Q(z)=\int_z^\infty \frac{1}{\sqrt{2\pi}}\exp\left(-\frac{t^2}{2}\right)dt$.
    Hence, the channel transition matrix $P$ is represented as a tridiagonal matrix. 
            
    Fig. \ref{fig:distribution} shows the threshold voltage distribution for $2$-bit per cell flash memories. 
    Within the constrained window $W$, there are four distinct states ranging from $s_1$ (erase state) to $s_4$ (highest state). 
    The window $W$ is defined as the distance between the mean of the erase state and the mean of the highest state, i.e., $W = \mu_{2^N} - \mu_1$.
    Once we determine the size-$(2^{N+1}-2)$ set $\{\Delta_{i,j}\}$, the distances between states are uniquely determined, and since the $\{\sigma_{i}\}$ parameters are assumed given, the $\{f_i(t)\}$ distributions are also determined. 

    \subsection{Channel-Aware Quantization} \label{subsec:caq}
    
    While the Lloyd-Max algorithm \cite{Lloyd1982least}, \cite{Max1960quantizing} aims to minimize the MSE, it does not account for channel errors. 
    In~\cite{Kurtenbach1969quantizing}, the authors proposed a channel-aware quantization technique, referred to as \emph{quantizing for noisy channels}, which improves the MSE by explicitly considering channel errors during the quantization process. 
    By accounting for channel errors, the MSE can be expressed as follows~\cite{Kurtenbach1969quantizing}:
    \begin{align}
    &\mathbb{E}[(X-\widehat{X})^2] \nonumber \\
    & = \int_{-\infty}^{\infty}x^2f_X(x) dx-2\sum_{j=1}^{M}v_j\sum_{i=1}^{M}P_{i,j} \nonumber \cdot \int_{u_i}^{u_{i+1}}x f_X(x)dx \nonumber \\
    & + \sum_{j=1}^{M}v_j^2\sum_{i=1}^{M}P_{i,j} \cdot \int_{u_i}^{u_{i+1}}f_X(x)dx. \label{eqn:quantization}
    \end{align}
    
    The channel-aware quantization aims to minimize~\eqref{eqn:quantization} by jointly optimizing $\{ v_j\}$ and $\{u_j\}$, as given in~\cite{Kurtenbach1969quantizing}:
    \begin{align}
    &v_j=\frac{\sum_{i=1}^M P_{i,j}\int_{u_i}^{u_{i+1}} x f_X(x)dx}{\sum_{i=1}^M P_{i,j}\int_{u_i}^{u_{i+1}}f_X(x)dx}, & j=1,\ldots,M, \label{eq:quant_value}
    \\ & u_j=\frac{1}{2} \cdot \frac{\sum_{k=1}^M v_k^2 (P_{j,k}-P_{j-1,k})}{\sum_{k=1}^M v_k(P_{j,k}-P_{j-1,k})}, & j=2,\ldots,M.\label{eq:quant_level}
    \end{align}
    As in the Lloyd-Max algorithm, the dependence between~\eqref{eq:quant_value} and~\eqref{eq:quant_level} requires some iterative algorithm to alternate between  \eqref{eq:quant_value} and \eqref{eq:quant_level} until convergence. 
    For a noiseless channel, the channel matrix simplifies to the identity matrix, i.e., $P_{i,j} = \delta_{i,j}$ (Kroneker delta). 
    Then, \eqref{eq:quant_value} and \eqref{eq:quant_level} reduce to the expressions for the Lloyd-Max algorithm.     
    
    \section{Integrated Optimization of Quantization Levels and Verify Levels}\label{sec:main}

    In this section, we propose an integrated framework that jointly optimizes quantization levels and verify levels to minimize the MSE of \eqref{eqn:quantization}. 
    The optimization variables include $ \{v_j\}$ and $\{u_j\} $ defining the quantization levels, and $\{\Delta_{i,j}\}$ determining the verify levels. 
    For simplicity, we set $\mathbf{u}:= \{u_j\}_{j=2}^{M}$, $\mathbf{v}:= \{v_j\}_{j=1}^{M}$, and $\mathbf{\Delta} := \{\Delta_{i,i+1},\Delta_{i+1,i}\}_{i=1}^{M-1}$. 
    Then, the joint optimization problem can be formulated as:
    \begin{align} \label{eq:opt_joint} 
    \underset{\mathbf{u}, \mathbf{v}, \mathbf{\Delta}}{\text{minimize}}~ ~&~ \text{MSE} (\mathbf{u},\mathbf{v},\mathbf{\Delta}) \nonumber  \\ 
    \text{subject to} ~&~ \| \mathbf{\Delta} \|_1 = W  \nonumber
    \\ 
    &  ~\mathbf{\Delta} \succeq 0, 
    \end{align}
    where $\| \cdot \|_1$ denotes the $\ell_1$-norm, and  $\mathbf{\Delta} \succeq 0$ implies $\Delta_{i,j} \ge 0$ for all $i$ and $j$.  

    This problem is non-convex, making it challenging to directly obtain the optimal solutions for $\mathbf{u}$, $\mathbf{v}$, and $\mathbf{\Delta}$. 
    Furthermore, the impacts of $\mathbf{u}$, $\mathbf{v}$, and $\mathbf{\Delta}$ on the MSE are interdependent. 
    Given a channel transition probability matrix, the optimal $\mathbf{u}$ and $\mathbf{v}$ can be determined iteratively using \eqref{eq:quant_value} and \eqref{eq:quant_level}, as described in Sec.~\ref{subsec:caq}. 
    However, $\mathbf{\Delta}$ determines the channel transition matrix, which in turn affects $\mathbf{u}$ and $\mathbf{v}$. 
    Conversely, the optimal $\mathbf{\Delta}$ also depends on $\mathbf{u}$ and $\mathbf{v}$ through the distribution $p(s_i)$ in~\eqref{eq:GMM}. 
    
    To address this interdependency, we propose an iterative algorithm to solve \eqref{eq:opt_joint}. 
    First, we initialize a starting point $\mathbf{\Delta}^{(0)}$ that satisfies the constraints of \eqref{eq:opt_joint}.
    Next, the corresponding channel transition matrix $P=[P_{i,j}]$ is computed, which can be approximated as a tridiagonal matrix using \eqref{eq:transition_prob_1} and \eqref{eq:transition_prob_2}. 
    Given $P=[P_{i,j}]$, the corresponding $\mathbf{u}$ and $\mathbf{v}$ are iteratively updated using \eqref{eq:quant_value} and \eqref{eq:quant_level}. 
    Subsequently, $\mathbf{\Delta}$ is optimized for the updated $\mathbf{u}$ and $\mathbf{v}$. 
    This process is then repeated iteratively. 

    The following proposition describes how to update $\mathbf{\Delta}$ for given $\mathbf{u}$ and $\mathbf{v}$.
    \begin{proposition}\label{prop:verify}
    For given $\mathbf{u}$ and $\mathbf{v}$, $\mathbf{\Delta}$ is updated by solving the following \emph{convex} optimization problem: 
    \begin{align} \label{eq:opt_verify}
    \underset{\mathbf{\Delta}}{\text{minimize}}~
    ~&~\sum_{i=1}^{M-1}\bigg\{ \gamma_{i,i+1} \cdot p(s_{i}) Q\left(\frac{\Delta_{i,i+1}}{\sigma_i}\right) \nonumber \\
    ~&~ +  \gamma_{i+1,i} \cdot p(s_{i+1})Q\left(\frac{\Delta_{i+1,i} }{\sigma_{i+1}}\right)  \bigg\}  \nonumber \\
    {\text{subject~to}} ~&~ \|\mathbf{\Delta}\|_1=W 
    \nonumber \\
    & ~\mathbf{\Delta} \succeq 0,
    \end{align}
    where $\gamma_{i,j}=\left( \widetilde{v}_i - v_j \right)^2$, $p(s_i) = \int_{u_i}^{u_{i+1}} f_X(x) dx$, and
    \begin{equation} \label{eq:v_tilde}
    \widetilde{v}_i = \frac{\int_{u_i}^{u_{i+1}}xf_X(x)dx}{\int_{u_i}^{u_{i+1}}f_X(x)dx}.
    \end{equation}
    \end{proposition}
    \begin{IEEEproof}
	The proof is given in Appendix \ref{pf:prop_verify}.
    \end{IEEEproof}
    Note that $p(s_i)$ and $\gamma_{i,j}$ are determined by the given $\mathbf{u}$ and $\mathbf{v}$.
    Since the optimization problem of \eqref{eq:opt_verify} is convex, it can be solved using standard methods~\cite{Boyd2004convex}. 

    The proposed iterative algorithm to jointly optimize $\mathbf{u}$, $\mathbf{v}$, and $\mathbf{\Delta}$ is summarized in Algorithm~\ref{algo:joint_opt}. 

    \begin{algorithm}
		\caption{Joint Optimization Algorithm for Quantization and Verify Levels} \label{algo:joint_opt}
		\begin{algorithmic}[1]
			\State Choose a starting point $\mathbf{\Delta}^{(0)}$ satisfying the constraints of \eqref{eq:opt_joint} and set $k=0$. 
			\Repeat
			\State Compute $P = [P_{i,j}]$ by using $\mathbf{\Delta}^{(k)}$. 
			\State Iteratively update $\mathbf{u}^{(k)}$ and $\mathbf{v}^{(k)}$ by \eqref{eq:quant_value} and \eqref{eq:quant_level}. 
			\State Update $\mathbf{\Delta}^{(k+1)}$ for given $\mathbf{u}^{(k)}$ and $\mathbf{v}^{(k)}$. \Comment{Prop.~\ref{prop:verify}}  
			\State $k := k+1$. 
			\Until{Stopping criterion is satisfied.}  
		\end{algorithmic}
	\end{algorithm}	
    \begin{figure}\centerline{\includegraphics[width=0.45\textwidth]{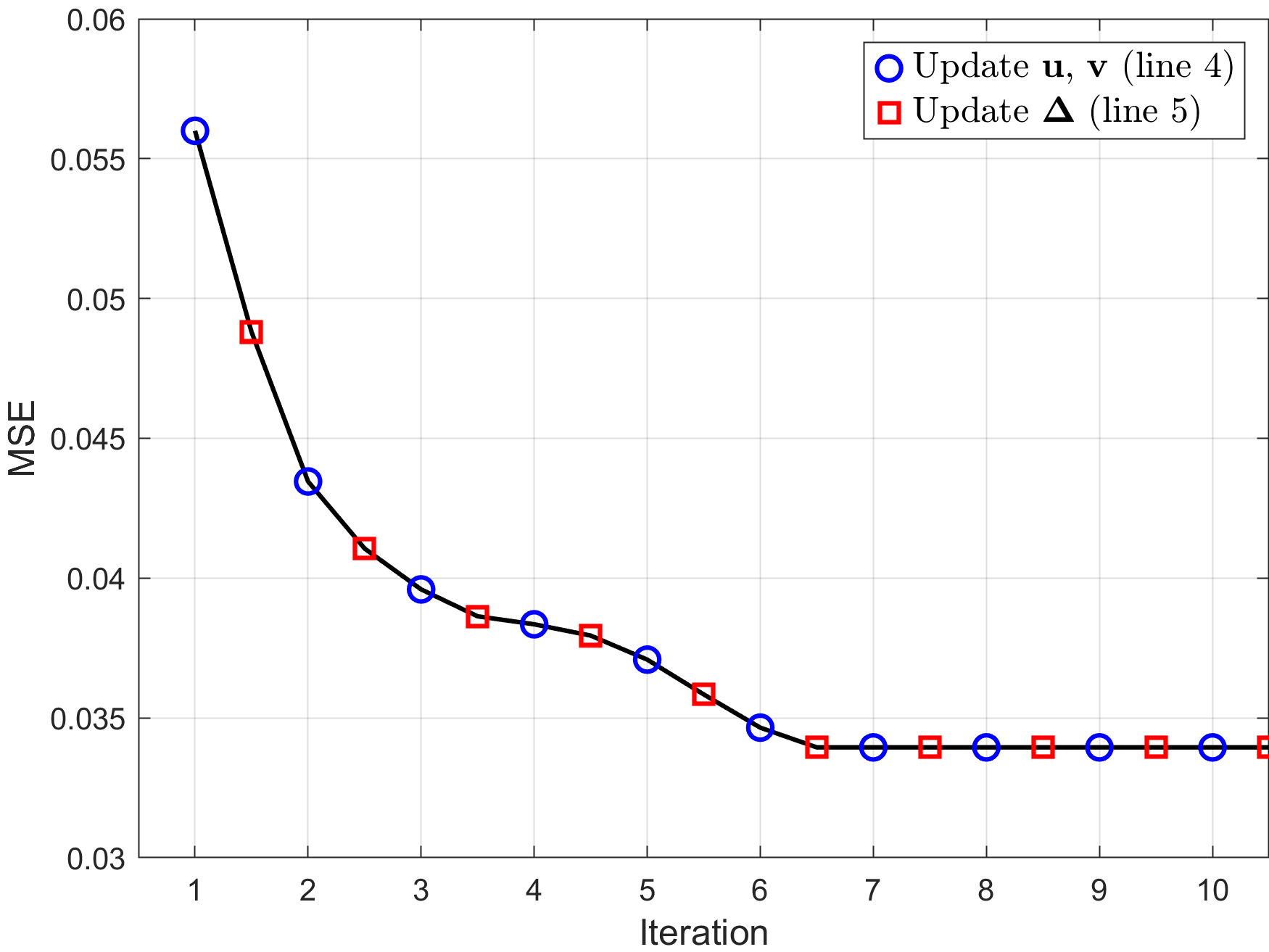}}
        \caption{Iterative joint optimization for $X \sim \mathcal{N}(0, 1)$, $\sigma=0.2$, $W=5$, and $M=16$.} 
        \label{fig:iteration} 
    \end{figure}

    \begin{figure}[t]
    \centerline{\includegraphics[width=0.45\textwidth]{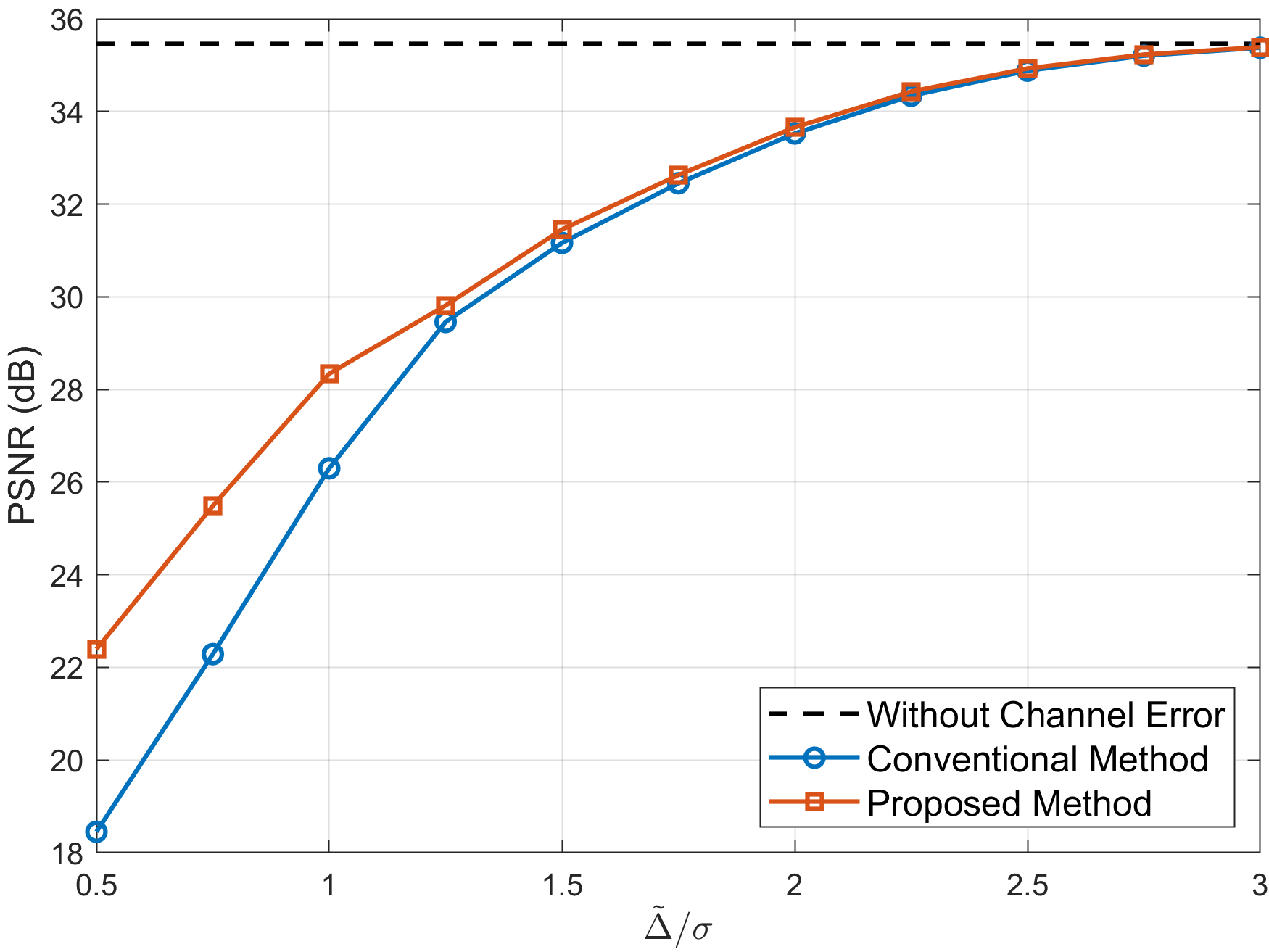}}
        \caption{Comparison of PSNRs for image retrieval.}
        \label{fig: BSD plot} 
    \end{figure}

    \begin{figure}[t]
    \captionsetup{font=small}
    \scriptsize
    \hspace{-0.3cm}
    \begin{tabular}{c@{\extracolsep{0.0em}}@{\extracolsep{0.0em}}c@{\extracolsep{0.0em}}c@{\extracolsep{0.0em}}}
    		\includegraphics[width=0.155\textwidth]{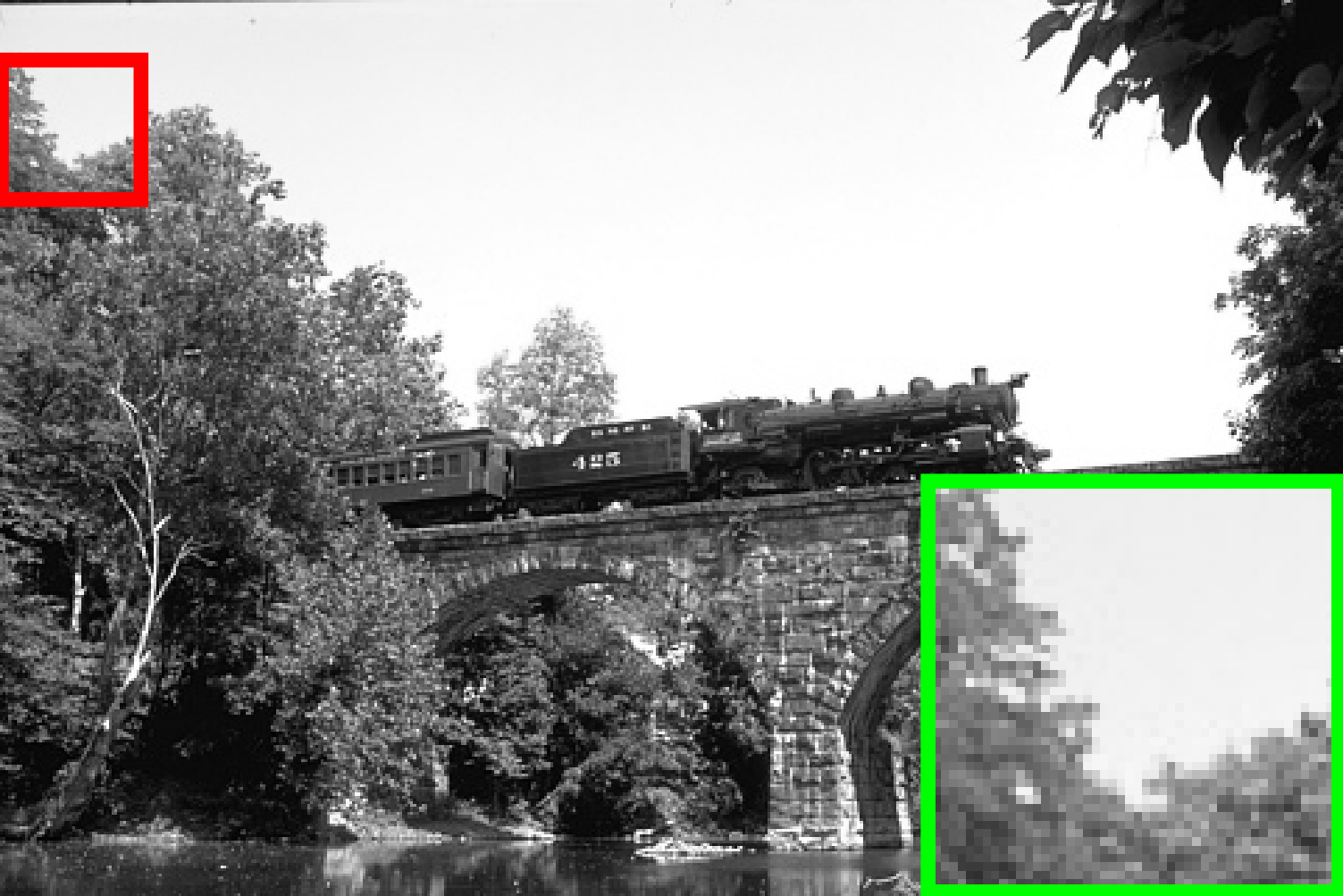}~~
    		&\includegraphics[width=0.155\textwidth]{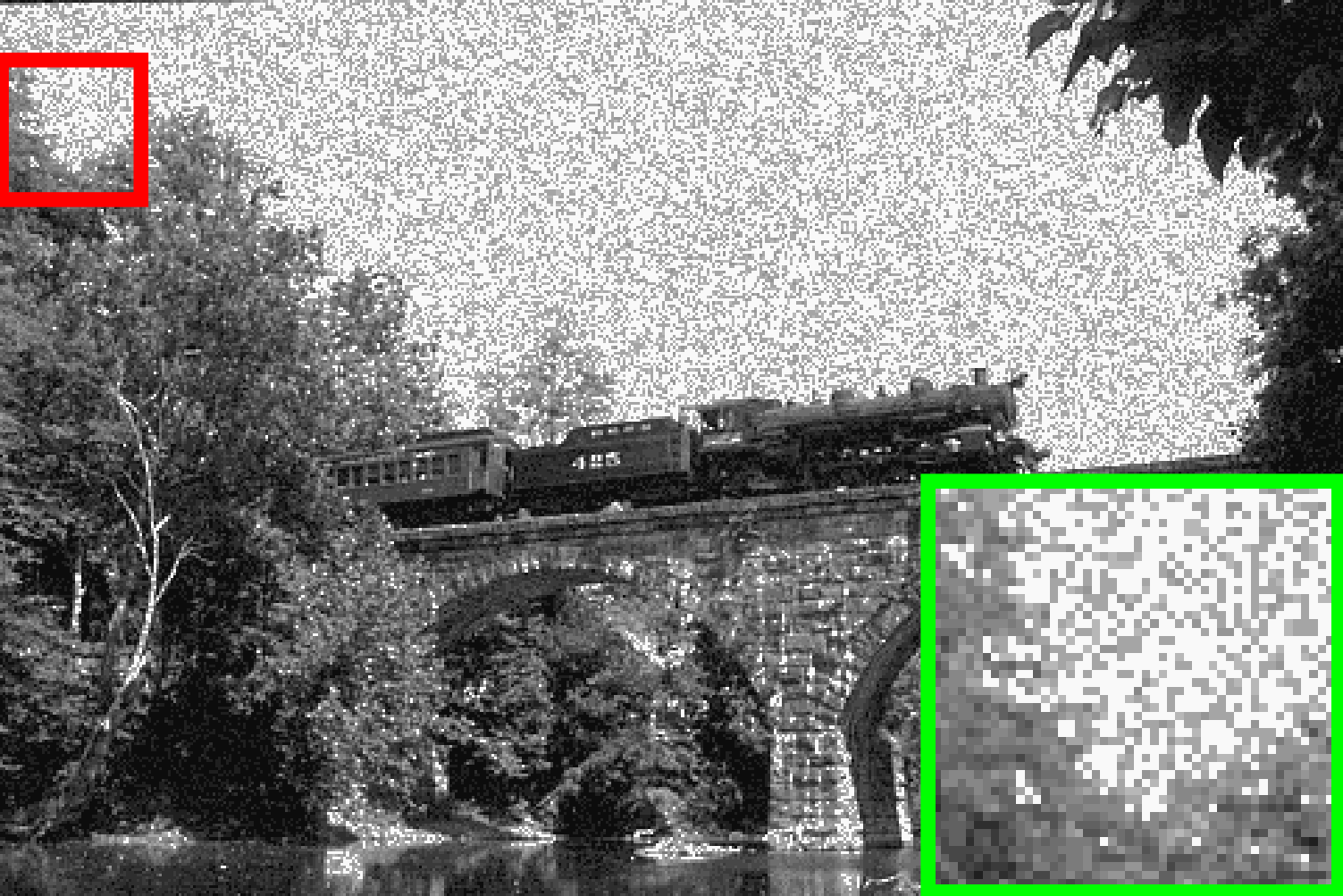}~~
    		&\includegraphics[width=0.155\textwidth]{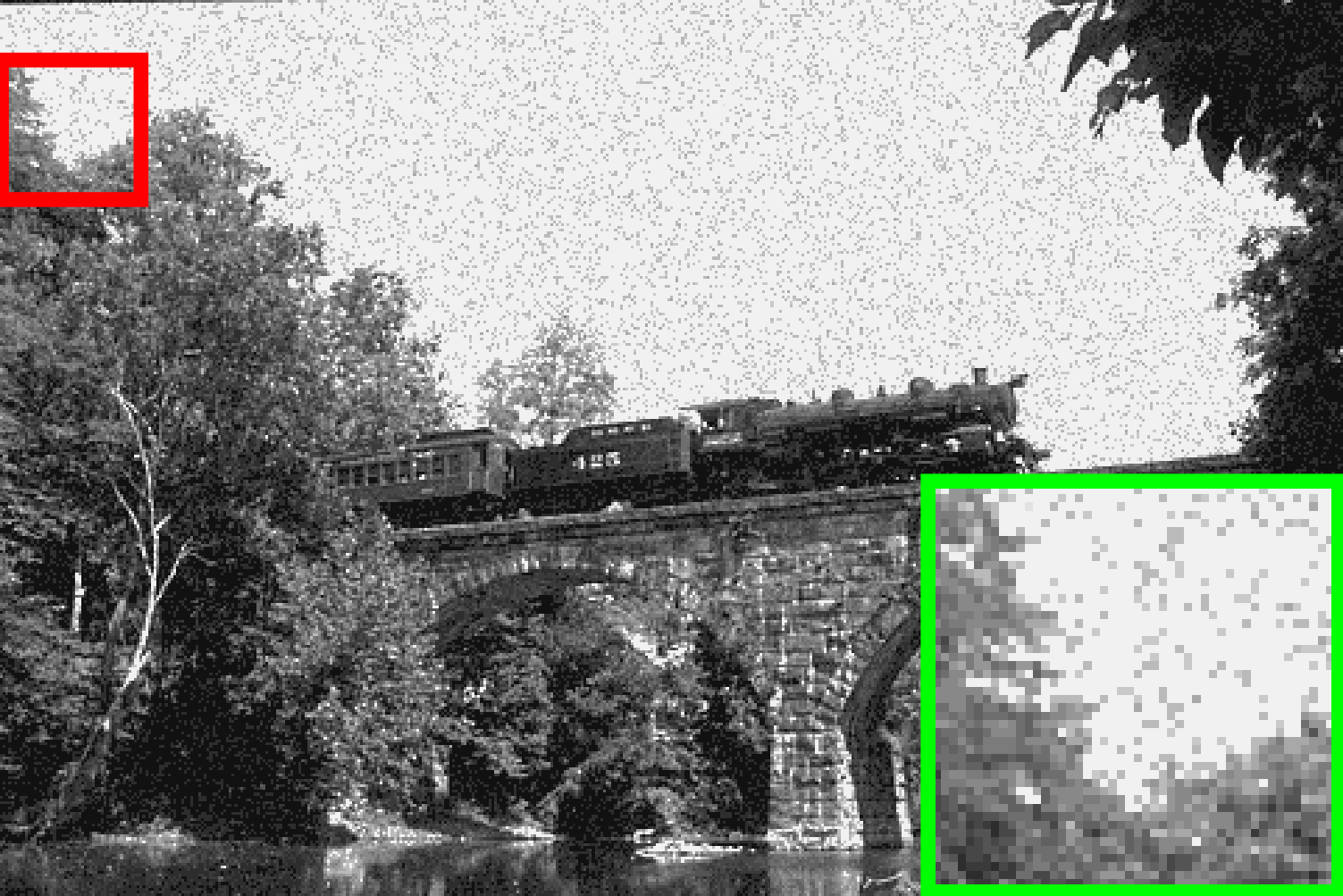}~~\\

     (a) & (b) & (c)\\
    \end{tabular}
    \caption{Visual comparison for image retrieval on the image ``\emph{test068}'' from BSD68 at $\widetilde{\Delta}/\sigma=0.75$: (a) original image, (b) conventional method (\SI{17.92}{dB}), and (c) proposed method (\SI{23.13}{dB}).}
    \label{fig:image_recovery}
    \end{figure}

    \begin{remark}
    Since \eqref{eq:opt_joint} is non-convex, a starting point can affect the final solutions.
    To address this, we begin by considering a noiseless channel where $P_{i,j}=\delta_{i,j}$, and determine $\mathbf{u}$ and $\mathbf{v}$ using the Lloyd-Max algorithm. 
    Next, we initialize a starting point $\mathbf{\Delta}^{(0)}$ that minimizes the overall BER. 
    Subsequently, we update $\mathbf{u}$, $\mathbf{v}$, and $\mathbf{\Delta}$ using Algorithm~\ref{algo:joint_opt}.
    \end{remark}
    
    \begin{remark} [Stopping Criterion]
    The stopping criterion can be defined in several ways. 
    For example, it may be based on the difference between $\text{MSE}(\mathbf{u}^{(k)},\mathbf{v}^{(k)},\mathbf{\Delta}^{(k)})$ and $\text{MSE}(\mathbf{u}^{(k+1)},\mathbf{v}^{(k+1)},\mathbf{\Delta}^{(k+1)})$, or the difference between $\mathbf{\Delta}^{(k)}$ and $\mathbf{\Delta}^{(k+1)}$. Alternatively, a maximum number of iterations can be used as the stopping criterion. 
    \end{remark}
    
    Fig. \ref{fig:iteration} shows the iterative reduction of the MSE over 10 iterations, eventually converging. 

    \section{Numerical Results}\label{sec:results}

    To evaluate the effectiveness of our proposed method, we assess its performance in image retrieval and 
    image super-resolution tasks by considering $4$-bit per cell quad-level cell (QLC) flash memory. 
    We compare the results of our proposed method with those of a conventional approach. 
    The conventional method consists of two separate steps: 
    1) quantization is performed without accounting for channel errors, which reduces to the solutions of the Lloyd-Max algorithm, 2) verify levels are optimized to minimize the overall BER as in~\cite{Kim2012verify} using the state probabilities obtained from the first step. 
    
    We define  $\widetilde{\Delta}=\frac{W}{2(M-1)}$, representing the average value of $\Delta$. 
    When solving \eqref{eq:opt_verify}, we assume uniform standard deviations across all states for simplicity.     
    However, the proposed algorithm is general and can accommodate arbitrary noise variances and window sizes. 
    
    \subsection{Quantized Image Retrieval}
    
    In the image retrieval experiments, we utilize the Berkeley Segmentation Dataset 68 (BSD68), a simplified version of the BSD100 \cite{Martin2002database} containing gray scale images.
    Each pixel of a grayscale image, represented as an 8-bit unsigned integer, is quantized to 4 bits and subsequently stored in an individual QLC flash cell.

    Fig. \ref{fig: BSD plot} compares the PSNRs of retrieved images using the proposed method with those restored by the conventional method. 
    The numerical results indicate that the proposed method significantly outperforms the conventional approach under higher noise variance, i.e., lower $\widetilde{\Delta}/{\sigma}$.
    For low noise variance, which is close to a noise-free channel, the PSNRs of the conventional method and the proposed method are nearly identical.  
    Additionally, the visual comparisons in Fig. \ref{fig:image_recovery} demonstrate that the proposed method produces noticeably fewer artifacts, resulting in improved overall perceptual quality.

    \subsection{Image Super-Resolution}

    \begin{figure}[t]
    \centering
    \subfloat[]{\includegraphics[width=0.45\textwidth]{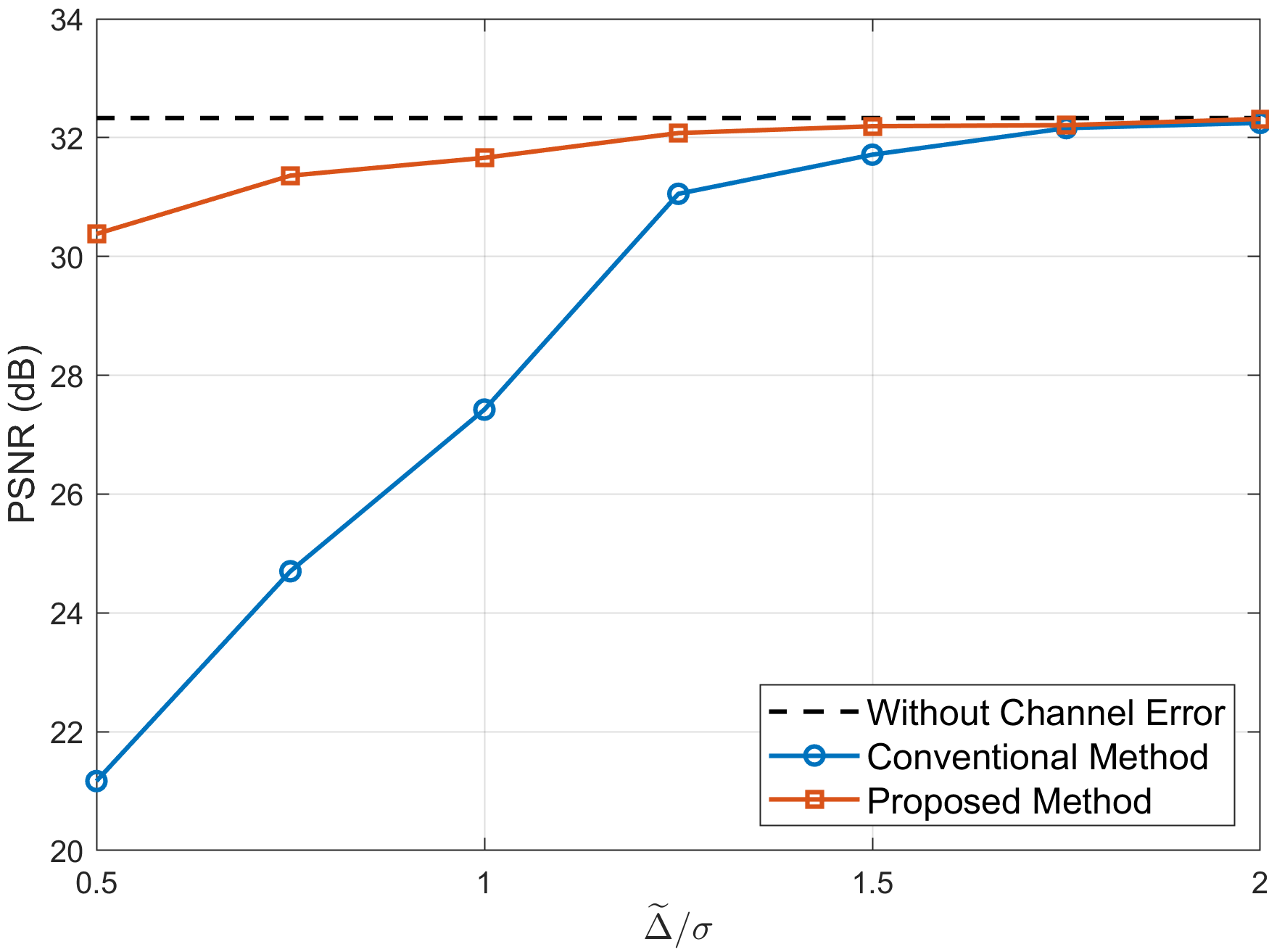}
    \label{fig:set5_case1}}
    \hfill
    \subfloat[]{\includegraphics[width=0.45\textwidth]{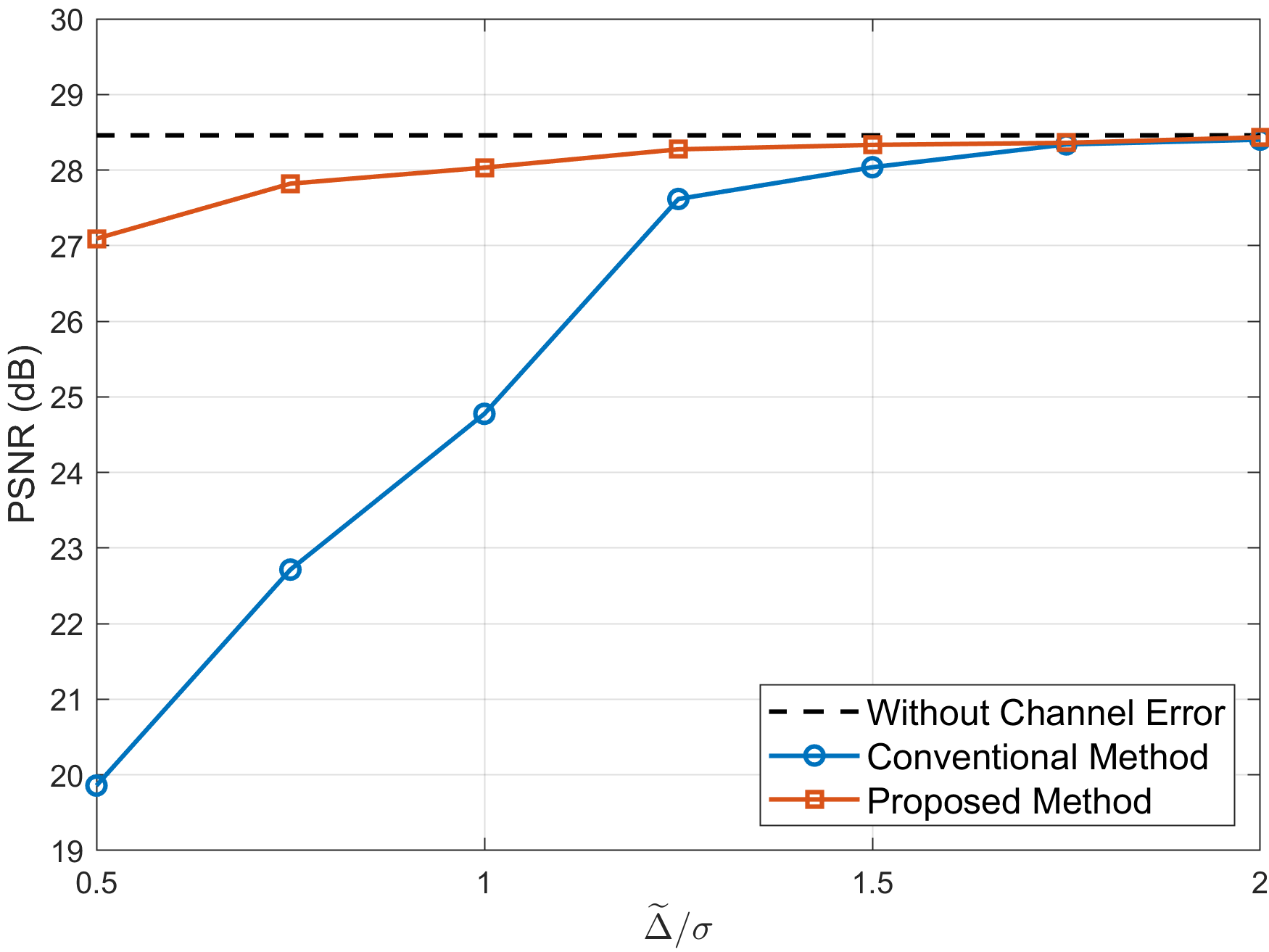}
        \label{fig:set14_case1}}
    \caption{Comparison of PSNRs for image super-resolution by SwinIR: (a) Set5 and (b) Set14.}
    \label{fig:PSNR_swinIR}
    \end{figure}

    \begin{figure}[t]
    \captionsetup{font=small}
    \scriptsize
    \hspace{-0.3cm}
    \begin{tabular}{c@{\extracolsep{0.0em}}@{\extracolsep{0.0em}}c@{\extracolsep{0.0em}}c@{\extracolsep{0.0em}}}
    		\includegraphics[width=0.155\textwidth]{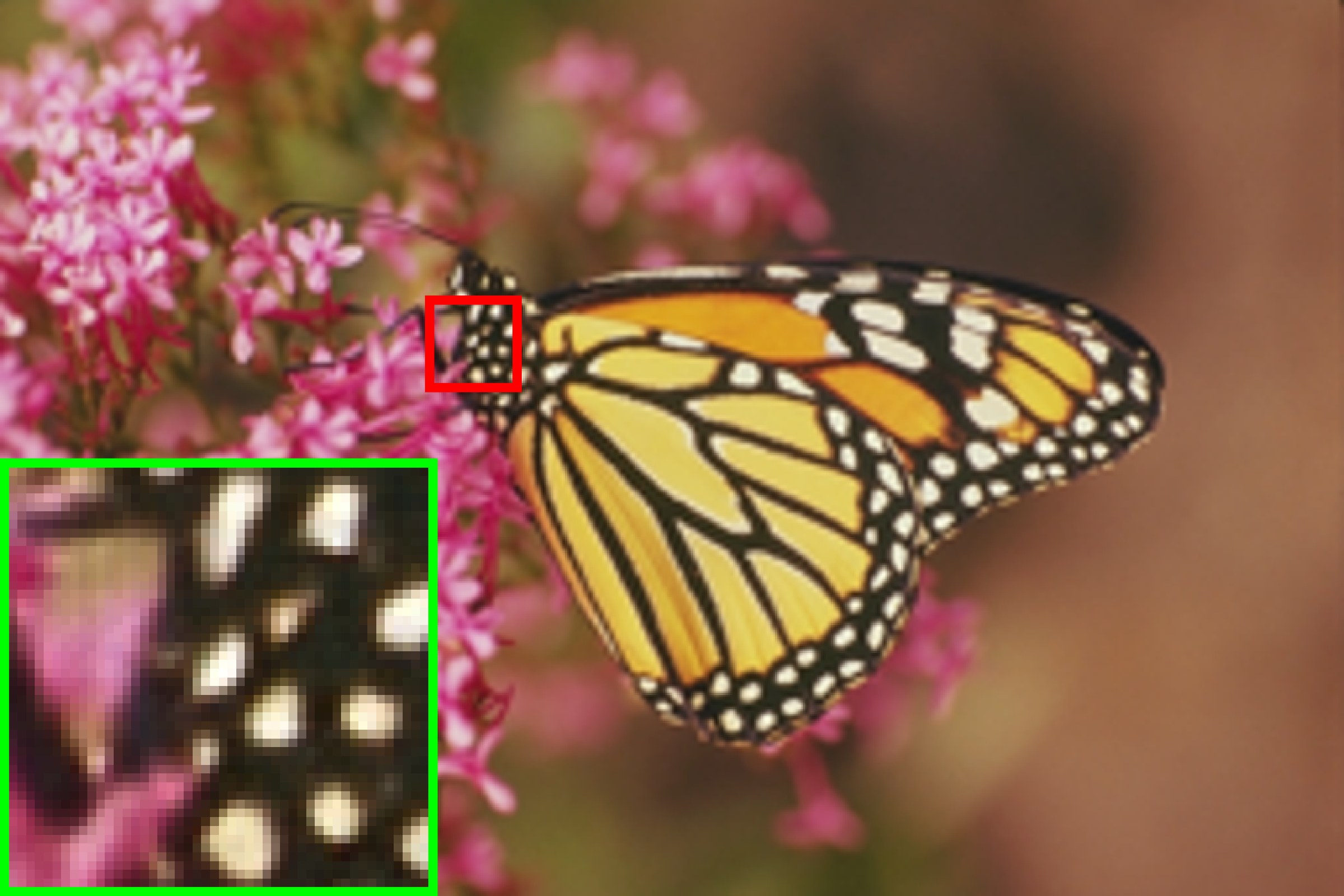}~~
    		&\includegraphics[width=0.155\textwidth]{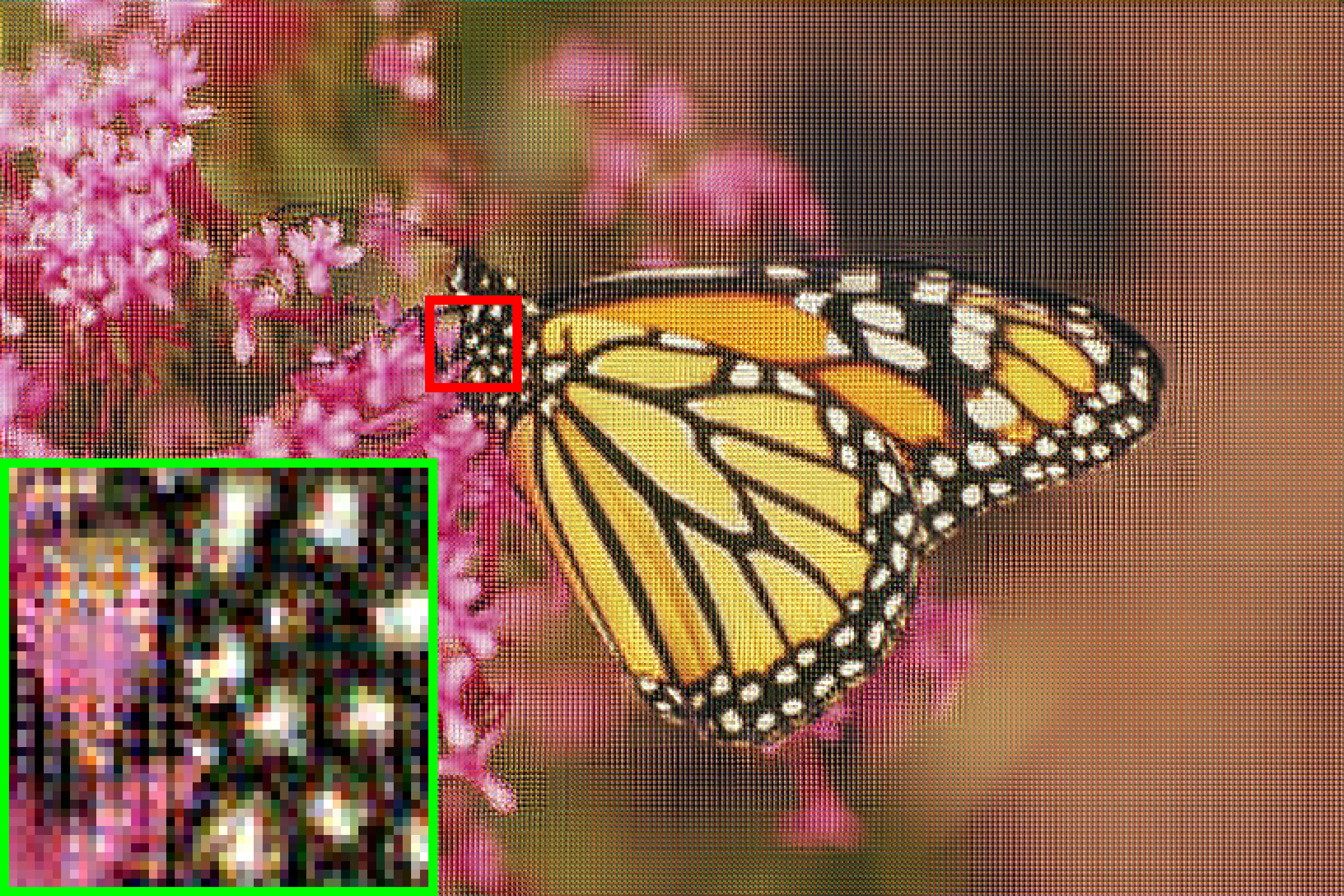}~~
    		&\includegraphics[width=0.155\textwidth]{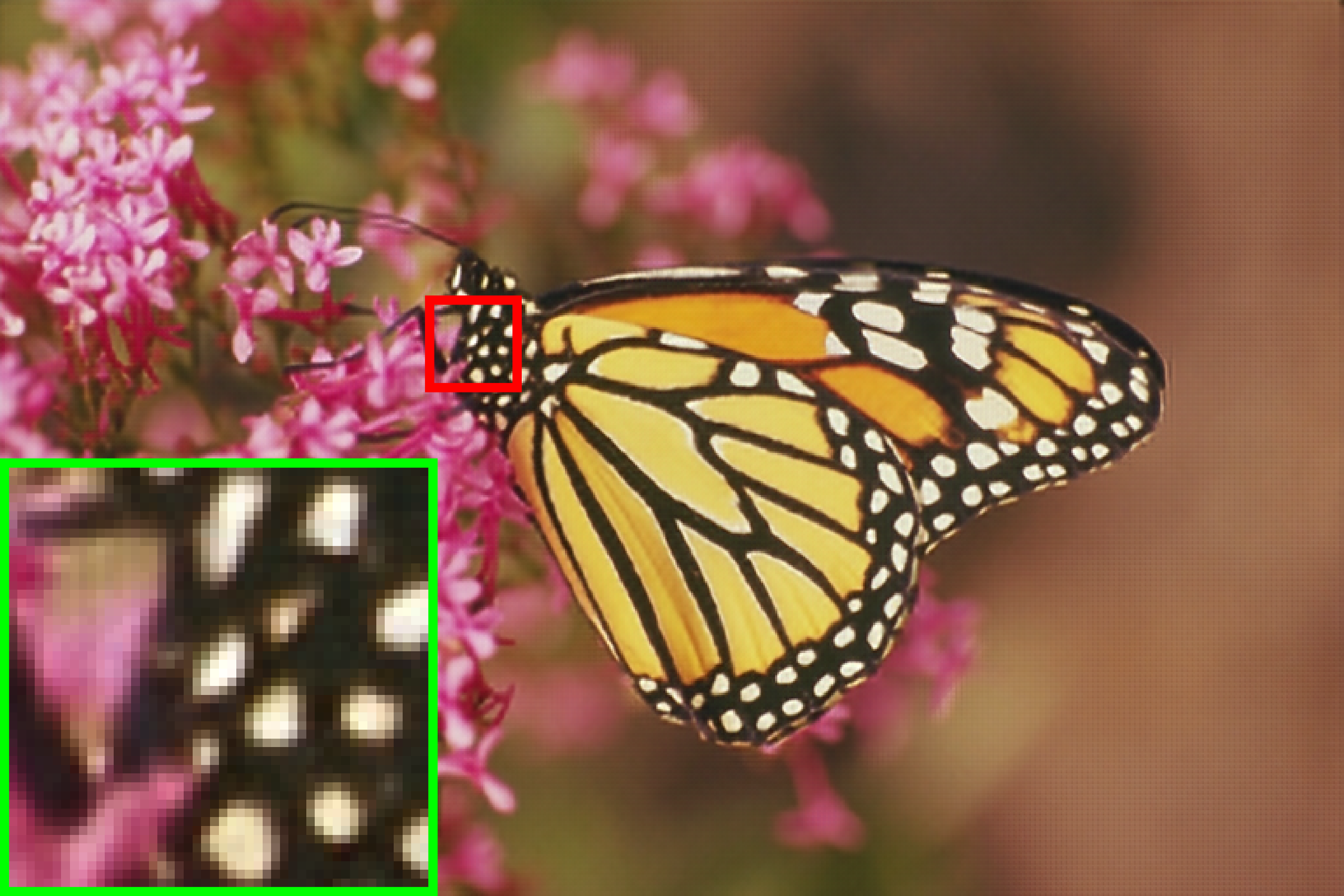}~~\\
     (a) & (b) & (c)\\
    \end{tabular}
    \caption{Visual comparison for the image super-resolution task on the ``\emph{monarch}'' image from Set14 at \(\widetilde{\Delta}/\sigma=0.5\): (a) low resolution, (b) conventional method (\SI{16.12}{dB}), and (c) proposed method (\SI{31.62}{dB}).}
    \label{fig:super_resolution}
    \end{figure}

    We employ the SwinIR-S$(\times 3)$ model \cite{Liang2021swinir} for the image super-resolution task on two datasets, Set5 \cite{Bevilacqua2012low} and Set14 \cite{Zeyde2012on}, commonly used for image quality evaluation.
    In the SwinIR model, the parameters are selectively quantized to $4$ bits, focusing on the multi-head self-attention (MSA) and multi-layer perception (MLP) sublayers.
    Specifically, the quantization is applied to the QKV projection, multi-head concatenation projection, and the two fully connected layers' parameters, as these modules account for the majority of the computational workload and parameter count in transformer-based architectures.
    Each quantized parameter is stored in a single QLC cell, with identical verify levels applied across all parameters.
    
    Fig. \ref{fig:PSNR_swinIR} compares the PSNRs of the SwinIR model for the conventional and proposed methods.
    The proposed method consistently outperforms the conventional approach, especially in the high noise variance range. 
    This indicates that the proposed method is robust to channel errors particularly in the high-error range.
    Fig. \ref{fig:super_resolution} presents a visual comparison of SwinIR for image super-resolution. 
    The results clearly show that the proposed method delivers sharper details, reduced artifacts, and higher-quality reconstructions, significantly outperforming the conventional method in super-resolution tasks.
    
    \section{Conclusion}\label{sec:conclusion}

    We proposed a novel integrated optimization of quantization and verify levels for flash-based PIM systems. 
    After formulating the joint optimization problem, we developed an iterative algorithm to improve the MSE under a given constraint. 
    Experimental results show that the proposed method effectively improves MSE performance for both image quantization and image super-resolution tasks.


\appendices
\section{Proof of Proposition~\ref{prop:verify}}\label{pf:prop_verify}

    The MSE can be expressed as follows:
    \begin{align}
    &\mathbb{E}_{X,\widehat{X}}[(X-\widehat{X})^2]  \nonumber \\
    &= \mathbb{E}_{Y,\widehat{X}}[\mathbb{E}_{X}[(X-\widehat{X})^2|Y,\widehat{X}]]  \nonumber \\
    \label{eqn:appendix1}
    &=\mathbb{E}_{Y,\widehat{X}}\left[\mathbb{E}_{X}[(X-\mathbb{E}[X|Y])^2|Y]+(\mathbb{E}[X|Y]-\widehat{X})^2\right] \nonumber \\
    &=\mathbb E_{X,Y}\left[(X-\mathbb E[X|Y])^2\right]\nonumber
    \\
    &\quad +\sum_{i=1}^{M}\sum_{j=1}^{M}p_{Y,\widehat X}(v_i,v_j)\left(\mathbb E[X|Y=v_i]-v_j\right)^2,
    \end{align}
    where $\mathbb E[X|Y=v_{i}] = \widetilde{v}_i$ by \eqref{eq:v_tilde}. Hence, $(\mathbb E[X|Y=v_i]-v_j)^2$ can be replaced with $\gamma_{i,j} = \left(\widetilde{v}_i-v_j\right)^2$ for $i,j\geq1$. 
    In \eqref{eqn:appendix1}, the first term is independent of optimizing the verify levels. 
    Consequently, we focus on the second term, which explicitly accounts for the channel errors. 
    The second term can be reformulated as follows:
    \begin{align}
    &\sum_{i=1}^{M}\sum_{j=1}^{M}p_{Y,\widehat X}(v_i,v_j)\left(\widetilde{v}_i - v_j\right)^2 \nonumber
    \\&=\sum_{i=1}^{M}p_Y(v_i)\sum_{j=1}^{M}p_{\widehat{X}|Y}(v_j|v_i)\left( \widetilde{v}_i - v_j\right)^2  \label{eqn:appendix 2}\\
    & \approx \sum_{i=1}^{M-1}     p_Y(v_{i}) P_{i,i+1} \cdot \gamma_{i,i+1}+p_Y(v_{i+1}) P_{i+1,i}\cdot \gamma_{i+1,i}   \label{eq:appendix_3}
    \\
    &\approx  
    \sum_{i=1}^{M-1}\bigg\{p(s_{i}) Q\left(\frac{\Delta_{i,i+1}}{\sigma_i}\right)\gamma_{i,i+1}  \nonumber \\
    & \quad +  p(s_{i+1})Q\left(\frac{\Delta_{i+1,i} }{\sigma_{i+1}}\right)\gamma_{i+1,i}  \bigg\} \label{eq:appendix_4},
    \end{align}
     where \eqref{eq:appendix_3} follows from the tridiagonal channel transition matrix, with the assumption that $\gamma_{i,i} \ll 1$. 
    Finally, \eqref{eq:appendix_4} follows from $p_Y(v_{i}) = p_S(s_i)$, \eqref{eq:transition_prob_1}, and \eqref{eq:transition_prob_2}.

    Note that \eqref{eq:appendix_4} is convex with respect to $\mathbf{\Delta}$ since $Q(\cdot)$ is a convex function and the coefficients are nonnegative.

    \section*{Acknowledgment}
    
    This work was supported by Institute of Information \& communications Technology Planning \& Evaluation (IITP) grant funded by the Korea Government (MSIT) (RS-2023-00229849, MPU/Connectivity/TinyML SoC solution for IoT intelligence with foundry-based eFLASH) and the National Research Foundation of Korea (NRF) grant funded by the Korean government (MSIT) (No. RS-2023-00212103).

    \IEEEtriggeratref{14}
    
    \bibliographystyle{IEEEtran}
    \bibliography{abrv,mybib}



\end{document}